\documentclass{PoS}
\DeclareRobustCommand{\ion}[2]{%
\relax\ifmmode
\ifx\testbx\f@series
{\mathbf{#1\,\mathsc{#2}}}\else
{\mathrm{#1\,\mathsc{#2}}}\fi
\else\textup{#1\,{\mdseries\textsc{#2}}}%
\fi}
\title{Low-column density HVC and IVC gas in the halo of the Milky Way}

\ShortTitle{Low-column density gas in the halo of the Milky Way}

\author{\speaker{N.~Ben Bekhti}$^a$, P.~Richter$^b$ \& M.~T.~Murphy$^c$\\
        \llap{$^a$}Argelander-Institut f\"{u}r Astronomie, University of Bonn\\
        \llap{$^b$}Institut f\"{u}r Physik und Astronomie, University of Potsdam\\
        \llap{$^c$}Centre for Astrophysics \& Supercomputing, Swinburne University of Technology\\
        E-mail: \email{nbekhti@astro.uni-bonn.de}}

\abstract{Recent studies of the circumgalactic gaseous environment of the Milky Way have concentrated on the distribution, chemical composition, and physical properties of the most massive neutral gas clouds and the highly-ionized halo absorbers. Relatively little effort has been put so far in exploring the circumgalactic neutral and weakly ionized metal absorbers at low HI column densities. 

With our work we systematically study the distribution and physical properties of neutral and ionised low-column density gas in the halo of the Milky Way. We combine CaII and NaI absorption line measurements with HI 21-cm emission line data. For some of the sight lines high-resolution radio synthesis observations were performed allowing us to study small-scale structures that cannot be resolved with single dish telescopes.

In total 177 lines of sight were observed, providing a large absorption-selected data sample for the analysis of IVC and HVC gas in the circumgalactic environment of the Milky Way. The study allows us to compare the observed absorption column density distribution (CDD) of gas in the Milky Way halo with the overall CDD of intervening absorbers towards quasars. The sensitive absorption line analysis enables us to identify the neutral and ionised gaseous structures at low column densities and small angular extent that possibly remain unseen in large 21-cm all-sky surveys. If this gas cover a significant portion of the sky, it possibly has a large influence on the evolution of the Milky Way.}

\FullConference{Panoramic Radio Astronomy: Wide-field 1-2 GHz research on galaxy evolution - PRA2009\\
		 June 02 - 05 2009\\
		 Groningen, the Netherlands}

\begin{document}

\section{Introduction}
High-resolution absorption and emission line measurements have 
demonstrated that the halo of the Milky Way contains a mixture of different gas phases. Multi-wavelength observations of this gaseous material provide insight into the various processes of the exchange of gaseous matter and energy between the galaxy and the intergalactic medium. 

A large fraction of the neutral gas in the halo is in the form of clouds with radial velocities inconsistent with a simple model of Galactic rotation, the so-called intermediate- and high-velocity clouds (IVCs, HVCs).
An efficient way to study metal abundances, small scale structures, and physical conditions in IVCs and HVCs is to combine absorption line measurements in the direction of quasars and HI 21-cm radio observations (Richter et~al. 2001 \cite{richref1}, Ben Bekhti et~al. 2009 \cite{benref1}).

Using UVES/VLT, we recently detected LVC, IVC, and HVC CaII and NaI absorption line systems in the direction of several quasars, suggesting the presence of gaseous substructures in the Galactic halo. Follow-up HI observations with the Effelsberg 100-m radio telescope have demonstrated that in many cases the CaII and NaI absorption is connected with neutral hydrogen gas with column densities of $10^{18} \ldots 10^{20}\,\mathrm{cm}^{-2}$. In some of these cases $N_\mathrm{HI}$ is below the 21-cm detection limit of large HI surveys (e.g., LAB). Additionally, four sight lines were observed  with high-resolution radio synthesis telescopes (WSRT and VLA) to search for small-scale structures embedded in the observed diffuse gas (Ben Bekhti et~al. 2009 \cite{benref2}).

%
%

\section{Results}

\begin{figure}[!t]
\centering
\includegraphics[width=0.9\textwidth]{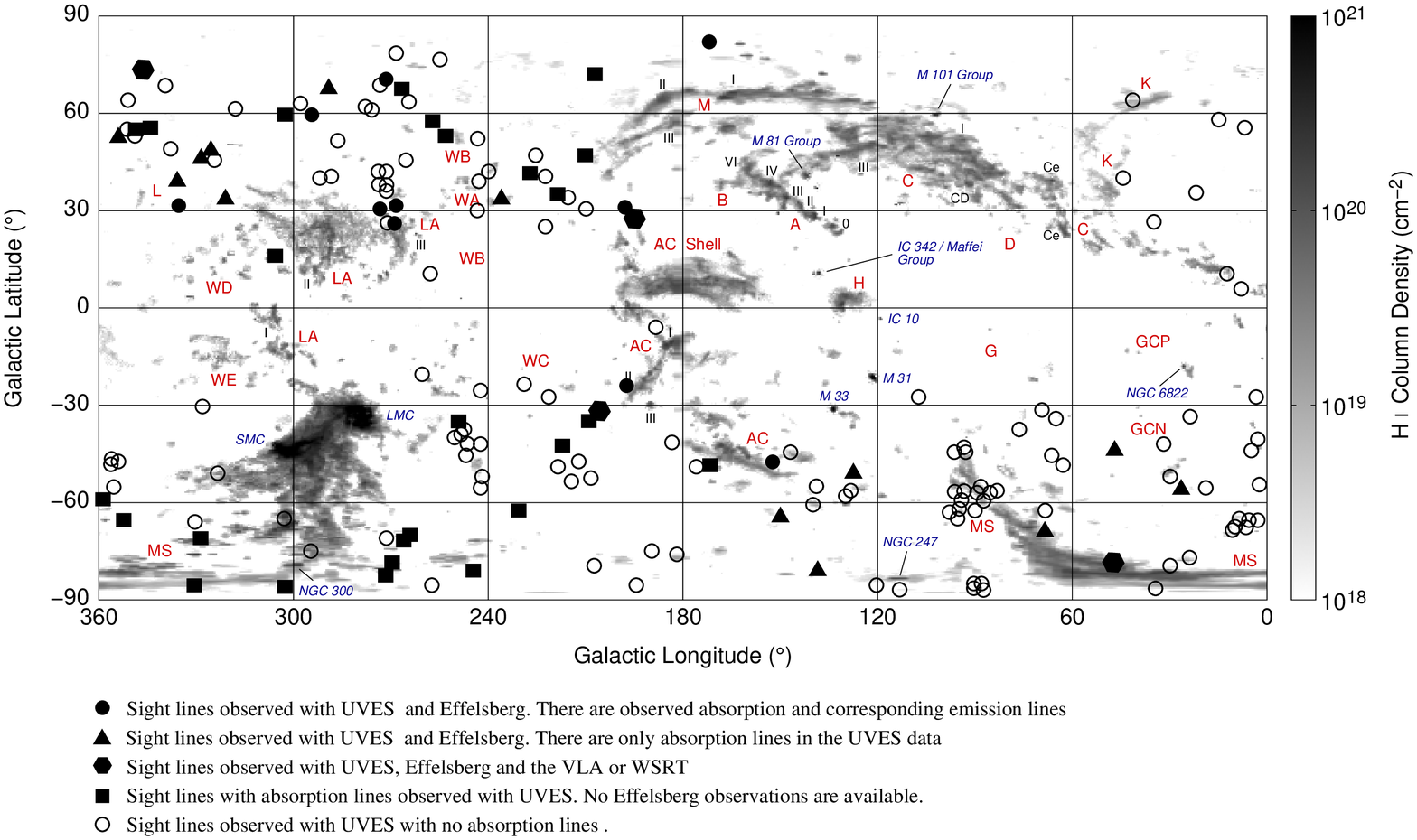}
\caption{All-sky HVC map derived from the LAB Survey (Kalberla et~al. 2005 \cite{kalberla05}). The different symbols mark the positions of 177~sight lines that were observed with UVES and 27~lines of sight where additional HI data were obtained with the 100-m telescope.}\label{fig_allsources}
\end{figure}

\begin{figure} 
\includegraphics[width=0.35\textwidth]{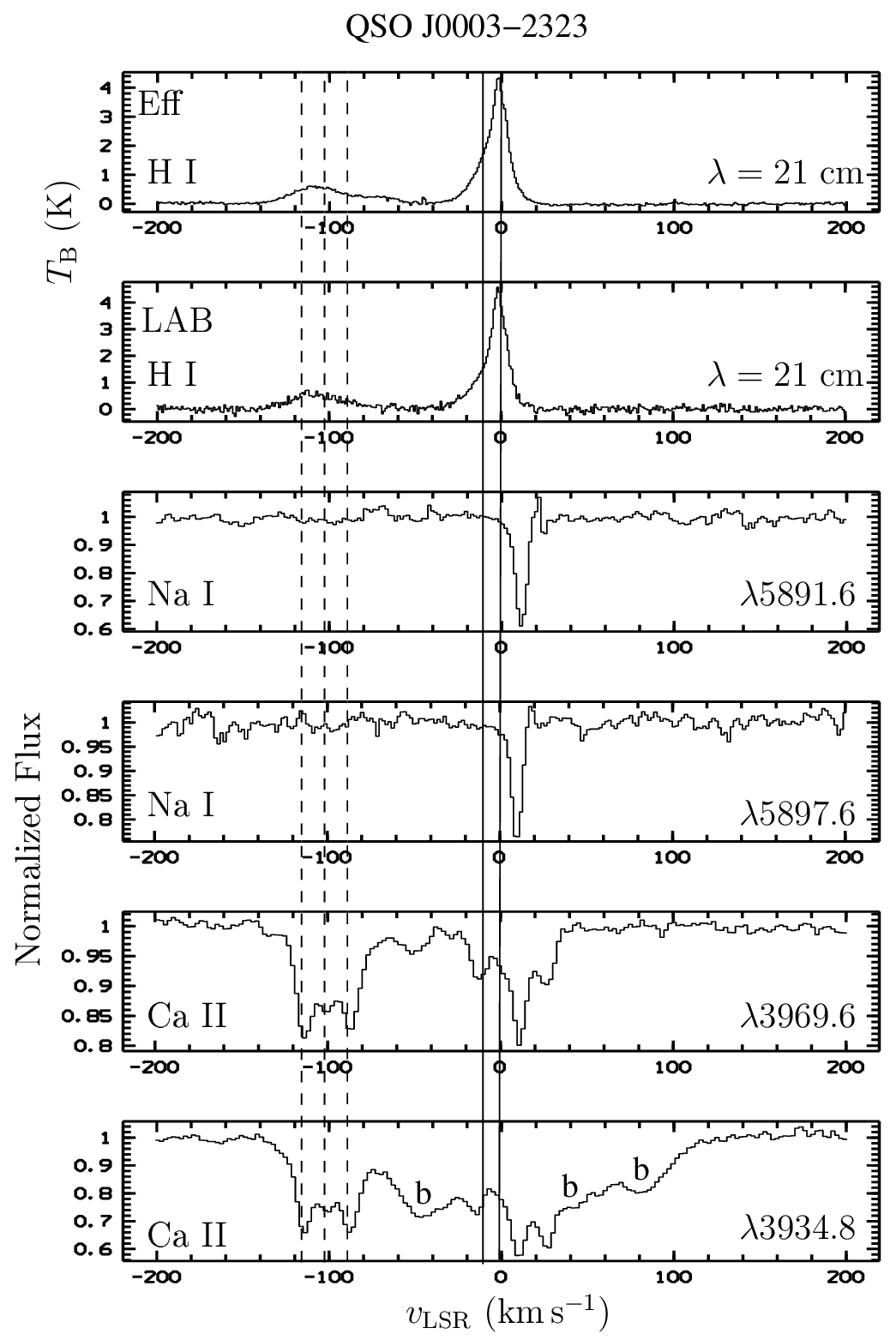}
\includegraphics[width=0.45\textwidth]{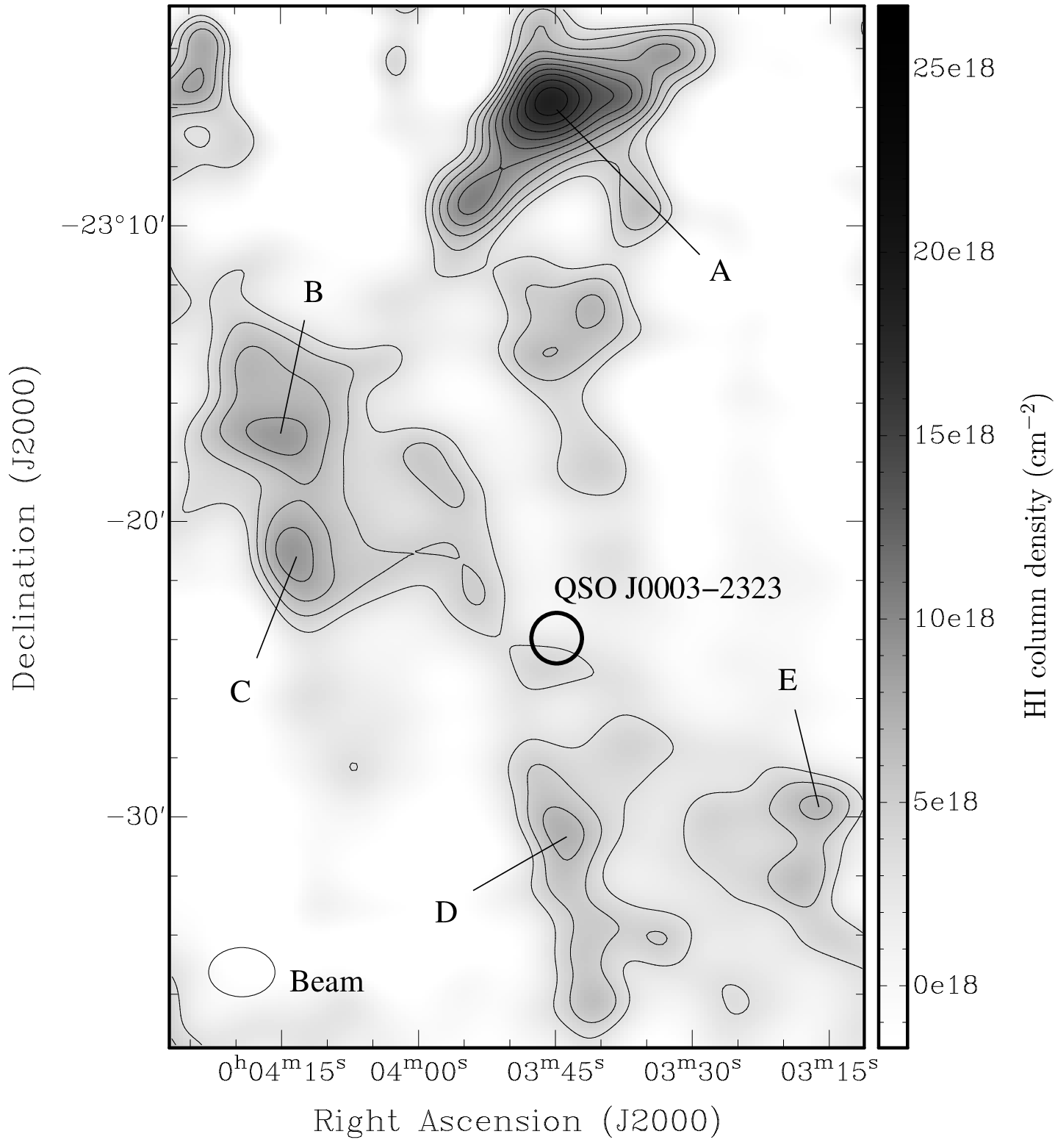} 
\caption{\textbf{Left}: Example spectra of an absorbing system towards QSO\,J0003$-$2323 observed with UVES, LAB, and Effelsberg. The absorption and emission lines are indicated by dashed lines. The solid lines mark the minimal and maximal LSR velocity which is predicted by a Milky Way model (Kalberla et~al. 2005 \cite{kalberla05}) for the galactic disc gas in this direction . \textbf{Right}: The corresponding high-resolution column density map observed with the VLA.} 
\label{fighighres} 
\end{figure}

Figure\,\ref{fig_allsources} shows an all-sky HVC map based on the LAB-survey (Kalberla et~al. 2005 \cite{kalberla05}). We find CaII (NaI) absorption in 47~(25)~out of 177~sight lines measured with UVES. Corresponding HI emission was detected with Effelsberg in 14~cases out of 27~measured lines of sight. Additionally, four sources were observed with high-resolution synthesis telescopes. The absorption/emission profiles of CaII, NaI, and HI for one exemplary system is shown in Fig.\,\ref{fighighres} together with the corresponding VLA column density map. The high-resolution observations reveal that cold ($70 \leq T_\mathrm{kin}^\mathrm{max} \leq 3700$\,K) and compact clumps are embedded in the diffuse IVC/HVC gas observed with Effelsberg, as previously reported for the source QSO\,B1448$-$232 (Richter et~al. 2005 \cite{richref2}). The fact that in the direction of 177~random sight lines we observe 53~absorption systems suggests that the Milky Way halo contains a large number of such low column density structures.

The CaII CDD follows a power-law with a slope of $\beta=-1.7$
(Ben Bekhti et~al. 2008 \cite{benref1}). This is similar to the result found for intervening MgII systems that trace the gaseous environment of other galaxies at low and high redshift (Churchill et~al. 2003 \cite{churchref2}). 


\begin{table*}
\caption{Parameters for four exemplary sight lines observed with UVES, and the possibly associated IVC/HVC complex.}
\small
\label{tab1}
\begin{tabular}{|l|c|c|c|c|c|c|}\hline
\rule{0pt}{3ex}Source&$l$\,[$^\circ$]&$b$\,[$^\circ$]&$v_\mathrm{LSR}$\,[km\,s$^{-1}$]&$N_\mathrm{CaII}$\,[cm$^{-2}$]&$b_\mathrm{CaII}$\,[cm$^{-2}$]&HVC/IVC compl.\\[1ex]\hline
\rule{0pt}{3ex}QSO\,J0003$-$2325&49.4&$-$78.6&$-$126&$7.7\cdot10^{11}$&6&MS \\\hline
\rule{0pt}{3ex}QSO\,B1331$+$170&348.5&75.8&$-$27&$1.0\cdot10^{12}$&5&IV Spur \\\hline
\rule{0pt}{3ex}QSO\,B0450$-$1310&211.8&$-$32.1&$-$20&$1.0\cdot10^{12}$&7&- \\\hline
\rule{0pt}{3ex}J081331$+$254503&196.9&28.6&$-$23&$1.3\cdot10^{11}$&5&- \\\hline
\end{tabular}
\end{table*}

\begin{table*}
\caption{Physical parameters for four clumps found with high-resolution observations.}
\small
\label{tab2}
\begin{tabular}{|l|c|c|c|c|c|c|}\hline
\rule{0pt}{3ex}Quasar&Clump&$v_\mathrm{LSR}$\,[km\,s$^{-1}$]&$b_\mathrm{HI}$\,[km\,s$^{-1}$]&$N_\mathrm{HI}$\,[cm$^{-2}$]&$T_\mathrm{kin}^\mathrm{max}$\,[K]&$\phi$\,[']\\[1ex]\hline
\rule{0pt}{3ex}QSO\,J0003$-$2325&A&$-$118&1.8&$1.9\cdot10^{19}$&200&3.1\\
\rule{0pt}{3ex}QSO\,B1331$+$170&A&$-$27&1.1&$1.2\cdot10^{19}$&70&1.8\\
\rule{0pt}{3ex}QSO\,B0450$-$1310&A&$-$18&4.6&$2.3\cdot10^{19}$&1250&2.7\\
\rule{0pt}{3ex}J081331$+$254503&A&$-$21&3.6&$2.0\cdot10^{19}$&800&2.7\\\hline
\end{tabular}
\end{table*}

About $40\%$ of the absorbers show multiple intermediate- and high-velocity components, indicating the presence of sub-structures. The direction, as well as the velocities of $50\%$ of the absorbers suggest an association with IVC/HVC complexes. Table\,\ref{tab1} summarises the physical parameters of the clouds which were observed with UVES, Effelsberg and WSRT/VLA. Table\,\ref{tab2} lists the parameters for the brightest clumps found for each of the sources. In the case of QSO\,J0003$-$2323 the HI single dish and interferometric data reveal only one emission line, while the absorption measurements exhibit a multi-component structure, showing that absorption spectroscopy is much more sensitive to low-column density gas. The radial velocities and spatial positions do not match exactly. It is possible that the different components and clumps trace quite different parts of the overall source. Such properties were already observed before (Br\"{u}ns et~al. 2001 \cite{bruensref1}) and confirmed simulations (Quilis \& Moore 2001 \cite{quilref1}), indicating interactions of the clouds with the ambient medium.

\end{document}